\documentclass[lettersize,journal]{IEEEtran}
\usepackage{amsmath,amsfonts}
\usepackage{algorithmic}
\usepackage{algorithm}
\usepackage{array}
\usepackage[caption=false,font=normalsize,labelfont=sf,textfont=sf]{subfig}
\usepackage{textcomp}
\usepackage{stfloats}
\usepackage{url}
\usepackage{verbatim}
\usepackage{graphicx}
\usepackage{multirow}  
\usepackage{cite}
\begin{document}

\title{ResPF: Residual Poisson Flow for Efficient and Physically Consistent Sparse-View CT Reconstruction}

\author{Changsheng Fang, Yongtong Liu, Bahareh Morovati, Shuo Han, Yu Shi, Li Zhou, Shuyi Fan, Hengyong Yu, ~\IEEEmembership{Fellow,~IEEE}
\thanks{ This work involved human subjects in its research. The authors confirm that all human subject research procedures and protocols are exempt from the Institutional Review Board at the University of Massachusetts Lowell (IRB\# 23-043). This work was supported in part by NIH/NIBIB under grants R01EB032807 and R01EB034737. } 
\thanks{The authors are with the Department of Electrical and Computer Engineering, University of Massachusetts Lowell, Lowell, MA 01854. 
H.Y. Yu serves as the corresponding author (email: hengyong-yu@ieee.org).}
}

\markboth{Journal of \LaTeX\ Class Files,~Vol.~14, No.~8, August~2021}%
{Shell \MakeLowercase{\textit{et al.}}: A Sample Article Using IEEEtran.cls for IEEE Journals}


\maketitle

\begin{abstract}
Sparse-view computed tomography (CT) is a practical solution to reduce radiation dose, but the resulting ill-posed inverse problem poses significant challenges for accurate image reconstruction. Although deep learning and diffusion-based methods have shown promising results, they often lack physical interpretability or suffer from high computational costs due to iterative sampling starting from random noise. Recent advances in generative modeling, particularly Poisson Flow Generative Models (PFGM), enable high-fidelity image synthesis by modeling the full data distribution. In this work, we propose Residual Poisson Flow (ResPF) Generative Models for efficient and accurate sparse-view CT reconstruction. Based on PFGM++, ResPF integrates conditional guidance from sparse measurements and employs a hijacking strategy to significantly reduce sampling cost by skipping redundant initial steps. However, skipping early stages can degrade reconstruction quality and introduce unrealistic structures. To address this, we embed a data-consistency into each iteration, ensuring fidelity to sparse-view measurements. Yet, PFGM sampling relies on a fixed ordinary differential equation (ODE) trajectory induced by electrostatic fields, which can be disrupted by step-wise data consistency, resulting in unstable or degraded reconstructions. Inspired by ResNet, we introduce a residual fusion module to linearly combine generative outputs with data-consistent reconstructions, effectively preserving trajectory continuity. To the best of our knowledge, this is the first application of Poisson flow models to sparse-view CT. Extensive experiments on synthetic and clinical datasets demonstrate that ResPF achieves superior reconstruction quality, faster inference, and stronger robustness compared to state-of-the-art iterative, learning-based, and diffusion models.
\end{abstract}


\section{Introduction}

Since the invention of X-ray computed tomography (CT) technology~\cite{cormack1963representation, hounsfield1973computerized}, it has become an essential tool for visualizing internal anatomical structures~\cite{wang2008outlook}, with widespread applications in clinical, industrial, and scientific fields~\cite{chen2017low}. With increasing use in diagnostic imaging, the associated radiation risks remain a major concern, because high-dose X-ray exposure is believed to increase the long-term risk of cancers~\cite{brenner2007computed}. Consequently, minimizing radiation dose while maintaining diagnostic image quality has become a critical research direction in CT system design and reconstruction algorithms~\cite{mk2004strategies}. According to the well-accepted principle ``As Low As Reasonably Achievable'' (ALARA) ~\cite{slovis2002alara}, two major dose reduction strategies have been proposed: (1) decrease X-ray intensity per projection angle~\cite{xu2012low, chen2014artifact}, and (2) reduce the number of projections when the system does not have a slip ring ~\cite{sidky2006accurate, chen2008prior, sidky2008image}, known as sparse-view data acquisition. The former is straightforward but susceptible to increased noise. The latter is more practical and efficient to improve temporal resolution but leads to incomplete projection data, resulting in severe streak artifacts~\cite{bian2010evaluation}, structural distortions, and even reconstruction failure~\cite{davison1983ill}.

This study focuses on the latter strategy: achieving high-quality image reconstruction from sparsely sampled CT projections with sufficient angular coverage. Existing methods for addressing this problem can be broadly categorized into three groups. Sinogram completion methods aim to restore missing projections before reconstruction, such as K-SVD-based dictionary learning~\cite{li2014dictionary, aharon2006k} and deep CNN-based interpolation with residual learning and patch-wise training~\cite{lee2017view, krizhevsky2012imagenet}. Iterative reconstruction (IR) methods combine statistical modeling in the projection domain with image-domain priors like sparsity~\cite{sidky2006accurate} or low-rank structures~\cite{cai2014cine}, including R-NLTV~\cite{kim2016non}, total generalized variation (TGV)~\cite{niu2014sparse}, and edge-preserving regularization~\cite{nien2015relaxed, kim2014combining}. Image post-processing methods first reconstruct an initial image using analytical techniques such as filtered backprojection (FBP), and then remove artifacts in image space, as in Siemens’ IRIS or the ALOHA method based on low-rank Hankel matrices~\cite{han2016sparse}.
Although these approaches have improved sparse-view CT reconstruction to different degrees, they face several challenges. Sinogram-based methods often rely heavily on distribution assumptions and struggle with severely undersampled data. Iterative approaches are computationally expensive and converge slowly, while post-processing techniques \cite{zhou2024rhonerfleveragingattenuationpriors} cannot fundamentally recover missing information. Moreover, most existing methods are \emph{discriminative models}, which focus on directly mapping from degraded to clean images, without modeling the data generation process. This makes it difficult to provide guarantees of physical consistency or distributional plausibility.

Recent advances in \emph{generative modeling} offer new opportunities for CT reconstruction. In particular, diffusion models and Poisson Flow Generative Models (PFGM) have demonstrated strong capabilities in modeling full data distributions and solving ill-posed inverse problems with high-fidelity image synthesis~\cite{liu2023diffusion,song2021solving, chung2022mr,ge2023jccs,han2024physics,morovati2024impact}. Diffusion models are inspired by non-equilibrium thermodynamics, using stochastic differential equations (SDE) or its deterministic ordinary differential equations (ODE) to gradually denoise Gaussian noise into structured images. EDM (Elucidating the Design Space of Diffusion-Based Generative Models)~\cite{karras2022elucidating} systematically improves noise scheduling and sampling efficiency. However, diffusion-based sampling typically requires hundreds to thousands steps, leading to high computational costs that limit practical deployment in real-time CT applications. To overcome this limitation, PFGM~\cite{xu2022poisson} was proposed as a novel paradigm based on electrostatics, embedding $N$-dimensional data in an $(N+D)$ dimensional spherical space, and tracking electric field lines under a Poisson potential to form deterministic sampling trajectories. Its enhanced variant, PFGM++~\cite{xu2023pfgm++}, theoretically proved that the sampling behavior converges to EDM when $D \to \infty$ and $r = \sigma \sqrt{D}$, making diffusion model a special case of PFGM++. Moreover, PFGM++ can directly inherit training and sampling mechanisms of EDM, improving efficiency while retaining the expressive power of generative models and supporting step-wise control and structural editing capabilities. 

However, existing diffusion and Poisson flow generative models primarily focus on unconditional image generation or single-modal tasks, and have not been systematically explored for complex conditional generation tasks like sparse-view CT, which involves a combination of imaging physics, measurement constraints, and image priors. Current methods often overlook the strong conditional dependence between measurements and image contents, and fail to enforce data consistency in a coherent and trajectory-preserving manner. In particular, inserting data-consistency projections at each sampling step often disrupts the continuous nature of the generative flow, leading to artifacts and unstable reconstructions.

To address the aforementioned challenges of sparse-view CT reconstruction, we propose a novel conditional generative framework — \textbf{Residual Poisson Flow (ResPF)}. Based on PFGM++, ResPF first extends it into a \textbf{conditional generative model (cPFGM)} by incorporating sparse/full-view training data paired to enable posterior modeling conditioned on partial measurements. To further accelerate inference, we introduce a \textbf{hijacking strategy} that skips the initial stages of random noise sampling, significantly reducing the number of sampling steps while maintaining high reconstruction quality. This shortcut may lead to degraded or unrealistic outputs due to insufficient diffusion. To overcome this issue, we embed a \textbf{data-consistency} into each sampling iteration, ensuring that the generated images remain faithful to the measured projections and comply with physical constraints. PFGM relies on a fixed-direction ODE determined by electrostatic field lines, and injecting data consistency at each step perturbs this trajectory, potentially causing instability and degraded generation quality. To address this conflict, we introduce a \textbf{residual fusion module}, inspired by ResNet, which linearly combines the generative output with its data-consistent correction. This strategy preserves the smooth evolution of the sampling path while guiding the reconstruction toward physically plausible and measurement-consistent solutions. 

The main contributions of this work are fourfold.

\begin{enumerate}
  \item For the first time, Poisson flow generative models are applied to sparse-view CT reconstruction, integrating generative modeling, fast sampling, and physical fidelity.
  
  \item The original PFGM++ is extended into a conditional generative model (cPFGM) to enable posterior sampling guided by sparse-view measurements. This cPFGM forms the sampling backbone of our proposed reconstruction pipeline.
  
  \item ResPF is proposed as a conditional PFGM++ framework that integrates hijacked sampling, data consistency, and residual fusion into a unified generative process.
  
  \item Extensive experimental results on simulation and clinical datasets show that our proposed ResPF achieves state-of-the-art performance on both distortion metrics and reconstruction quality while offering noticeable speedup compared to previous diffusion-based methods.
\end{enumerate}

\section{Preliminary}
\subsection{Problem Formulation}

\begin{figure}[!t]
\centerline{\includegraphics[width=\columnwidth]{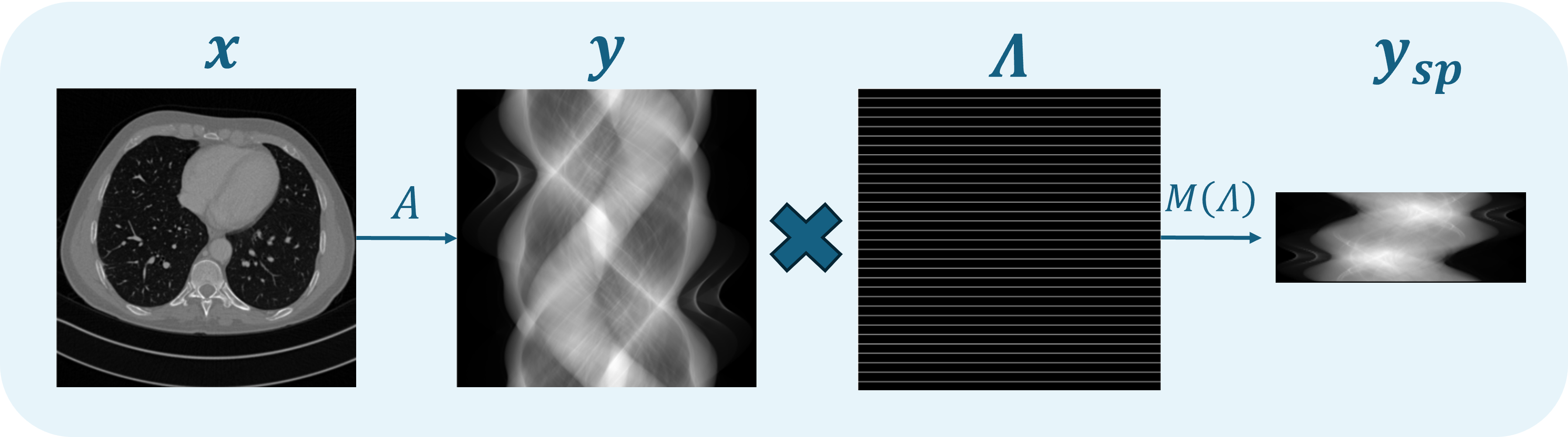}}
\caption{Modeling of sparse-view CT data acquisition.}
\label{fig:defination}
\end{figure}

In computed tomography (CT) image reconstruction, our goal is to reconstruct an image \(\mathbf{x} \in \mathbb{R}^N\) from projection measurements \(\mathbf{y} \in \mathbb{R}^M\). It can be modeled by the following linear system:
\begin{equation}
\mathbf{y} = \mathcal{A} \mathbf{x} + \boldsymbol{\eta},
\end{equation}
where \(\mathcal{A} \in \mathbb{R}^{M \times N}\) denotes the full-view system projection operator, and \(\boldsymbol{\eta}\) represents the post-log noise term. 
In an ideal scenario where full-scan projections \(\mathbf{y}\) are available, traditional analytic methods (\emph{e.g.}, filtered backprojection, FBP) can be used to reconstruct the image. However, in practical scenarios, due to constraints such as radiation dose, safety, or scanning time, sparse-view projections are acquired. The sparse sampling process can be modeled as:
\begin{equation}
\mathbf{y}_\text{sp} = \mathcal{M}(\Lambda) (\mathcal{A} \mathbf{x}+\boldsymbol{\eta}),
\end{equation}
where \(\mathcal{M}(\Lambda)\) denotes a sampling operator defined by the angular sub-sampling mask \(\Lambda\), and \(\mathbf{y}_\text{sp}\) represents the sparse-view projection measurements. The modeling process is illustrated in Fig.\ref{fig:defination}. Because the acquisition angles are sparse, directly applying methods like FBP often results in severe artifacts (\emph{e.g.}, streaking), degrading image quality. To recover high-quality images from \(\mathbf{y}_\text{sp}\), the problem is often formulated as a regularized inverse problem:
\begin{equation}
\min_{\mathbf{x}} \frac{1}{2} \|\mathcal{M}(\Lambda)\mathcal{A}\mathbf{x} - \mathbf{y}_\text{sp}\|_2^2 + \omega \mathcal{R}(\mathbf{x}),
\end{equation}
where \(\mathcal{R}(\mathbf{x})\) represents a prior regularization term (\emph{e.g.}, total variation, dictionary sparsity, deep priors), and \(\omega\) balances data consistency and prior strength. However, in extremely sparse or noisy cases, this formulation often fails to recover complete structures or fine details, thereby motivating the need for strong generative priors.

\subsection{EDMs}
EDM \cite{karras2022elucidating} provides a new perspective by integrating diffusion models into a unified framework for modular design. This not only enhances the quality and efficiency of image synthesis but also significantly increases the flexibility and scalability of the models. Compared to earlier models, it can synthesize high-quality images with fewer sampling steps. This design allows us to study and improve individual components of the model in a targeted manner, thereby more effectively exploring the feasible design space.

\subsubsection{Training}

EDM adopts a denoising-based training paradigm similar to that of conventional score-based diffusion models, but incorporate refined noise scheduling and loss weighting to improve convergence and sample quality. The training objective is defined as:

\begin{equation}
\mathbb{E}_{p(y),\,p(\sigma)} \left[ \lambda(\sigma) \left\| f_\theta(y + \sigma \epsilon;\, \sigma) - y \right\|_2^2 \right].
\label{edm_equ}
\end{equation}
The strategic selection of these distributions and functions plays a crucial role in substantially improving the overall performance of the diffusion model. Here, \( f_\theta \) denotes the denoising network parameterized by \( \theta \), trained to recover the clean data \( y \) from its noisy counterpart \( y + \sigma \epsilon \),  \( p(y) \) refers to the distribution of the actual data, \( p(\sigma) \) signifies the distribution from which the sampling of \( \sigma \) is derived, \( \lambda(\sigma) \) is the weighting function applied to the loss, and \( \epsilon \) represents the stochastic noise. 

This loss formulation allows the network to learn a family of denoising functions across a continuum of noise levels. Notably, the output of \( f_\theta \) is not the score (\emph{i.e.}, gradient of log-density), but rather the predicted clean signal. This enables direct integration into ODE-based samplers, such as DDIM or EDM samplers, by approximating the reverse-time trajectory via denoising estimates. The careful design of \( p(\sigma) \) and \( \lambda(\sigma) \) is crucial, as it governs the trade-off between reconstruction accuracy at low noise levels and model robustness at high noise levels.

\subsubsection{ODE Curvature Scheduling}
Through ODE curvature scheduling, the EDM uncovers a close relationship between lower ODE trajectory curvature and the demands for sampling steps. As the curvature decreases, the consistency in the direction of the tangent increases, which helps to reduce truncation errors when larger sampling steps are applied. Based on this discovery, EDM has designed a noise schedule that employs a simplified ODE process. In the selection of the noise schedule and sampling method, EDM focuses on achieving minimal curvature in the ODE trajectory, thus improving sampling efficiency and accuracy.

\subsubsection{Sampling}
EDM introduces an enhanced sampling technique that incorporates Heun's method \cite{ascher1998computer} along with a pioneering approach to time-step selection. This refined methodology significantly minimizes truncation errors and facilitates the creation of high-fidelity images with an optimized number of sampling steps and significantly enhances FID performance across multiple popular datasets. Specifically, let \(f_\theta(z_t, t)\) denote the estimator at time step \(t\) for the data sample \(z\), where \(\theta\) represents a pre-trained deep neural network model. The score function \(V_z \log q(z_t; \alpha_t, \sigma_t) = -\frac{(z_t - \alpha_t f_\theta(z_t,t))^2}{2\sigma_t^2}\) can then be reformulated at the given time step \(t_i\) to facilitate the computation of the next diffusion state \(z_{i-1}\).
The selection of time-step is governed by a specific equation:

\begin{equation}
t_i = \left( \sigma_{\text{max}}^\frac{1}{\rho} + \frac{i}{n-1} (\sigma_{\text{min}}^\frac{1}{\rho} - \sigma_{\text{max}}^\frac{1}{\rho}) \right)^\rho,
\end{equation}
where \( n \) denotes the total number of steps, and \( t_i \) is the time-step chosen at the \( i^{th} \) step. The parameter \( \rho \) is utilized to modulate the stride length. A higher value of \( \rho \) indicates broader strides near the maximum noise level \( \sigma_{\text{max}} \) and narrower strides near the minimum noise level \( \sigma_{\text{min}} \), thereby enhancing the image generation process. In the case of EDM, the value of \( \rho \) is set to 7. Considering Heun's method as a second-order differential equation solver, the number of function evaluations (NFE) during the sampling process with EDM is \( 2n - 1 \).
Using this integrated approach, EDM not only reduces the number of sampling steps while maintaining image quality, but also optimizes the efficiency and accuracy of the
entire image generation process through precise control of the
ODE trajectory’s curvature.

\subsection{Poisson Flow Generative Model Framework}

\begin{figure}[!t]
\centerline{\includegraphics[width=0.75\columnwidth]{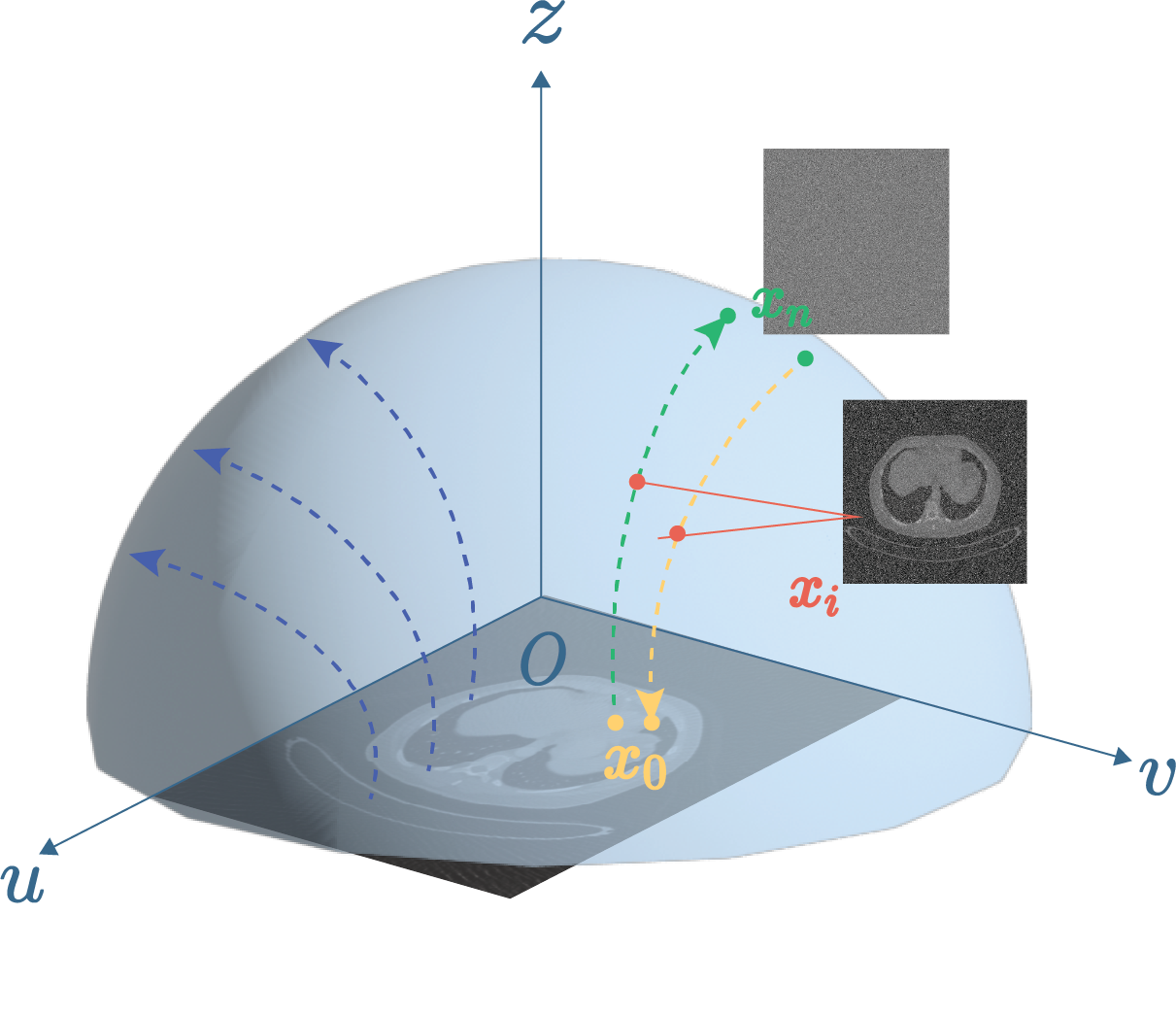}}
\caption{Illustration of 3D Poisson field trajectories for a 2D CT image distribution. The evolvements of a distribution or an augmented sample through the forward/backward ODEs pertains to the Poisson field.}
\label{fig:pfgm}
\end{figure}

Instead of estimating a time-dependent score function like score-based diffusion models, PFGM++ focuses on high-dimensional electric field to augment the \textit{N}-dimensional data to \textit{N+D}-dimensional space (see Fig. \ref{fig:pfgm} for a \textit{3D} Poisson field, here \(N=2, D=1\)):
\begin{equation}
\mathbf{E}(\tilde{u}) = \frac{1}{S_{N+D-1}(1)} \int \frac{\tilde{u} - \tilde{v}}{\|\tilde{u} - \tilde{v}\|^{N+D}} p(v) \, d\tilde{v},
\end{equation}
where \(p(v)\) is the ground truth data distribution, \(S_{N+D-1}(1)\) is the surface area of the unit \((N+D-1)\)-sphere, \(\tilde{v} := (v, 0) \in \mathbb{R}^{N+D}\), and \(\tilde{u} := (u, z) \in \mathbb{R}^{N+D}\) represent augmented data. In this framework, data points are treated as electric charges in augmented space, establishing a surjection between the ground truth data distribution and a uniform distribution on the infinite \((N+D)\)-dimensional hemisphere. The electric field's rotational symmetry on the \(D\)-dimensional cylinder \(\sum_{i=1}^{D} z_i^2 = r^2\) allows a reduction in dimensionality. By tracking \(r = r(\tilde{u})=\|\mathbf{z}\|_2\), the augmented variables are redefined as \(\tilde{v} := (v, 0) \in \mathbb{R}^{N+1}\) and \(\tilde{u} := (u, r) \in \mathbb{R}^{N+1}\). Here, the perturbation radius $r$ in high-dimensional space is a key parameter that guides data points from the initial noise distribution to the target distribution. Along the electric field line trajectories, the interest ODE is:
\begin{equation}
\frac{du}{dr} = \frac{\mathbf{E}(\tilde{u})_u}{\mathbf{E}(\tilde{u})_r},
\end{equation}
where, $\mathbf{E}(\tilde{u})_u = \frac{1}{S_{N+D-1}(1)} \int \frac{u - v}{\|\tilde{u} - \tilde{v}\|^{N+D}} p(v) \, dv, \quad$ $\mathbf{E}(\tilde{u})_r = \frac{1}{S_{N+D-1}(1)} \int \frac{r}{\|\tilde{u} - \tilde{v}\|^{N+D}} p(v) \, dv$.
This reduction establishes a bijection between data on the \(r = 0\) hyperplane (\(z = 0\)) and a distribution on the \(r = r_\text{max}\) hyper-cylinder. PFGM++ adopts a perturbation objective similar to the denoising score-matching objective in diffusion models. Using a perturbation kernel \(p_r(u|v)\), the objective is:
\begin{equation}
\label{equ:loss pfgm}
\mathbb{E}_{r \sim p(r)} \mathbb{E}_{v \sim p(v)} \mathbb{E}_{u \sim p_r(u|v)} \left\| f_\theta(\tilde{u}) - \frac{u - v}{r / \sqrt{D}} \right\|_2^2,
\end{equation}
where \(p(r)\) is the training distribution over \(r\). By choosing perturbation kernel \(p_r(u|v) \propto (||u - v||_2^2 + r^2)^{(N+D)/2}\), the minimizer of this objective aligns with the ODE is $f_\theta^*(\tilde{u}) = \sqrt{D} \, \frac{\mathbf{E}(\tilde{u})_u}{\mathbf{E}(\tilde{u})_r}.$ Starting from an initial sample from \(p_{r_\text{max}}\), samples from the target distribution can be generated by solving the ODE solver numerically.

\section{Methodology}
\begin{figure*}[!t]  
  \centering
  \includegraphics[width=0.9\textwidth]{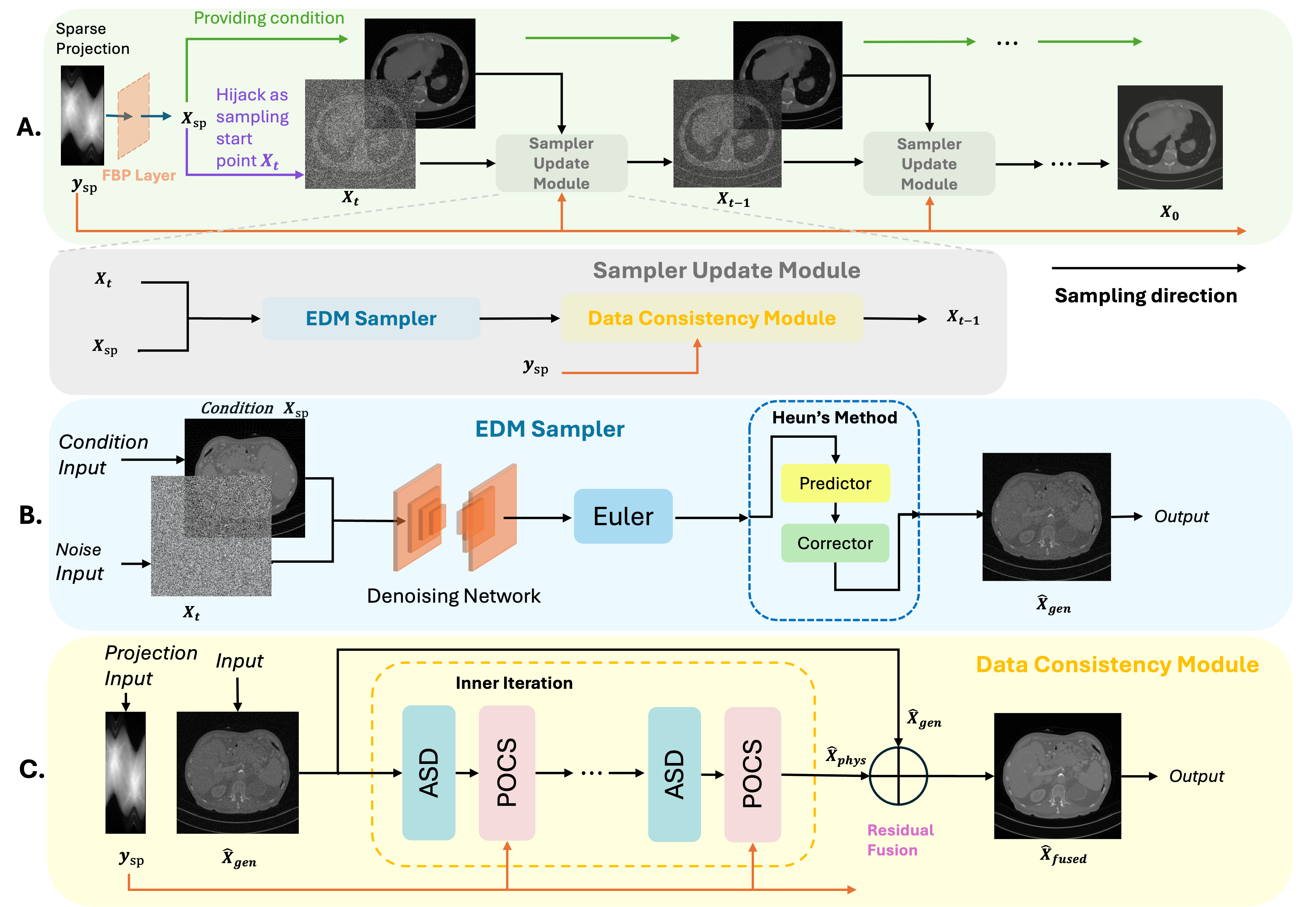}
  \caption{
Sampling process of the proposed Residual Poisson Flow (ResPF) model. 
(A) A sparse-view reconstruction $\mathbf{x}_{\text{sp}}$ is used both to hijack the sampling at timestep $t$ and as a conditioning input throughout the sampling chain. 
(B) The EDM sampler updates samples using noise and condition inputs. 
(C) Data consistency is enforced via ASD-POCS, and residual fusion yields the next sample for iterative refinement. More details can be seen in Algorithm \ref{alg:respf_sample}.
}

\label{fig:flow chart}
  
\end{figure*}

Fig.\ref{fig:flow chart}(A) illustrates the sampling process of our proposed Residual Poisson Flow (ResPF) framework. It integrates conditional generation, physics-informed correction, and residual fusion into a unified generative process. By jointly leveraging structural priors and measurement consistency, ResPF enables high-quality CT reconstruction even under extremely sparse-view conditions.
The complete sampling procedure is detailed in Algorithm \ref{alg:respf_sample}, which outlines the interplay between the hijacking strategy, data consistency updates, and residual fusion. Moreover, the proposed fusion strategy is easy to implement and flexible, allowing seamless adaptation to different noise levels and sampling densities. It functions as a tunable post-processing mechanism that enhances both the perceptual quality and diagnostic reliability of the reconstructed images.

\subsection{Optimized Conditional Poisson Flow Generative Model (cPFGM)}
The Poisson Flow Generative Model (PFGM) is a class of non-Markovian generative models based on potential field modeling. Its key idea is to construct an electric field in an extended high-dimensional space, which is induced by the data distribution. By solving the Poisson equation, the resulting field direction guides particles to evolve from a Gaussian prior distribution toward the true data distribution. Compared to traditional diffusion models, PFGM offers significant advantages in modeling global structural consistency and controlling generation trajectories.

In the sparse-view CT reconstruction task, our goal is to recover a high-quality full-view image \(\hat{\mathbf{x}}\) from undersampled observations. This process can be formalized as a conditional generation problem:
\begin{equation}
\hat{\mathbf{x}} \sim p(\mathbf{x} \mid \mathbf{x}_\text{sp}),
\end{equation}
where \(\mathbf{x}_\text{sp}\) is an initially reconstructed image obtained from sparse-view projections using FBP. According to Bayes’ theorem, the posterior distribution can be decomposed as:

\begin{equation}
    p(\mathbf{x} \mid \mathbf{x}_\text{sp}) \propto p(\mathbf{x}_\text{sp} \mid \mathbf{x}) \cdot p(\mathbf{x}),
\end{equation}
where \(p(\mathbf{x})\) is the prior distribution of natural CT images, and \(p(\mathbf{x}_\text{sp} \mid \mathbf{x})\) corresponds to the degradation process from a full-view image to a sparse-view reconstruction.

To more effectively guide the generative process, we explicitly incorporate the observation information into the PFGM framework as a condition to create a Conditional Poisson Flow Generative Model (cPFGM). The cPFGM serves as the foundation for our full model but should be distinguished from the proposed ResPF framework. 
While it is theoretically feasible to constrain the generative model directly on projection data (\emph{e.g.}, sinograms), we find that using low-quality images, reconstructed by standard FBP or regularized least squares (RLS), as conditional inputs significantly simplifies model training. This strategy has been widely adopted in traditional CT reconstruction literature \cite{han2018framing,liu2021sgd, jin2017deep}. We choose FBP to reconstruct conditional images due to its high computational efficiency, simplicity, and ability to preserve basic structural information while providing sufficient guidance for the generative model. Finally, we use FBP image \(\mathbf{x}_\text{sp}\) as the conditional input to the generative model.

During training, we sample data \(\mathbf{x}_0\) from the distribution of real images and compute the corresponding FBP conditional image \(\mathbf{x}_\text{sp}\). Then, a noise scale \(\sigma\) is sampled from a standard distribution, and the sampling radius is computed as \(r = \sigma \sqrt{D}\), where \(D\) is the extended dimension.
To construct the perturbed state, we sample a Gaussian direction vector \(\boldsymbol{\epsilon} \sim \mathcal{N}(0, I)\) in the high-dimensional space, and normalize it to obtain a spherical perturbation:
\begin{equation}
    \mathbf{x}_t = \mathbf{x}_0 + r \cdot \frac{\boldsymbol{\epsilon}}{\|\boldsymbol{\epsilon}\|}.
\end{equation}
We design a conditional neural network \(f_\theta\), whose input consists of the perturbed image \(\mathbf{x}_t\), the conditional image \(\mathbf{x}_\text{sp}\), and the noise scale \(\sigma\), aiming to predict the direction from the perturbed point toward the true image \(\mathbf{x}_0\). The training objective is defined as below from Eq.(\ref{edm_equ}):
\begin{equation}
     \mathcal{L}(\theta) = \mathbb{E}_{\mathbf{x}_0, \mathbf{x}_\text{sp}, \sigma, \boldsymbol{\epsilon}} (\lambda(\sigma))\left\| f_\theta(\mathbf{x}_t, \mathbf{x}_\text{sp}, \sigma) - \mathbf{x}_0 \right\|_2^2.   
\end{equation}
During training, we concatenate the perturbed image \(\mathbf{x}_t\) and the conditional image \(\mathbf{x}_\text{sp}\) along the channel dimension as the input to the network, thus establishing a conditional guidance mechanism that directs the sampling trajectory to gradually converge to structurally consistent regions.

\subsection{Conditional Hijacking Step}

Traditional Poisson flow or diffusion-based generative models typically initialize sampling from a Gaussian prior and iteratively refine the sample through hundreds of steps, gradually evolving from a random noise state to a clean image. While this ``from-scratch'' strategy is for general purpose, it has high inference costs and limits the applicability of such models in real-world scenarios like medical image reconstruction.
To address this issue, we propose a Conditional Hijacking Strategy, which significantly shorten the sampling trajectory and improve efficiency while maintaining sample quality. The key idea is to intervene directly in the generative trajectory using the conditional image, rather than starting from random noise. Specifically, we initialize sampling from an intermediate timestep $t_0$, thereby skipping the early high-noise stages. Relative EDM sampler is shown in Fig.\ref{fig:flow chart}(B).

\noindent{\emph{Initial State Construction}:}  
We use the conditional image \(\mathbf{x}_\text{sp}\) as the sampling starting point and perturb it to obtain an intermediate state:
\begin{equation}
\mathbf{x}_{t_0} = \mathbf{x}_\text{sp} + \sigma(t_0) \cdot \boldsymbol{\epsilon}, \quad \boldsymbol{\epsilon} \sim \mathcal{N}(0, I),
\end{equation}
where \(\sigma(t_0)\) denotes the noise level at timestep \(t_0\). In the EDM framework, this operation is interpreted as ``hijacking'' the trajectory at a later stage, constructing a structurally aligned initialization for generation.

\noindent{\emph{Sampling Update}: }  
As shown in Fig.\ref{fig:flow chart}(B), We adopt Heun's method to integrate the Poisson-driven trajectory. Heun's method is a second-order explicit predictor-corrector scheme. For each step \(t_i \to t_{i+1}\), it proceeds as follows:
\begin{align}
\mathbf{d}_i &= \hat{\mathbf{E}}_\theta(\mathbf{x}_{t_i}, \mathbf{x}_\text{sp}, t_i), \\
\tilde{\mathbf{x}}_{t_{i+1}} &= \mathbf{x}_{t_i} + (t_{i+1} - t_i) \cdot \mathbf{d}_i, \quad \text{(predictor)} \\
\tilde{\mathbf{d}}_{i+1} &= \hat{\mathbf{E}}_\theta(\tilde{\mathbf{x}}_{t_{i+1}}, \mathbf{x}_\text{sp}, t_{i+1}), \\
\mathbf{x}_{t_{i+1}} &= \mathbf{x}_{t_i} + \frac{1}{2}(t_{i+1} - t_i)(\mathbf{d}_i + \tilde{\mathbf{d}}_{i+1}). \quad \text{(corrector)}
\end{align}
Here, \(\hat{\mathbf{E}}_\theta\) denotes the model-predicted Poisson field direction (\emph{i.e.}, normalized gradient), and \(\mathbf{x}_\text{sp}\) is the conditional input concatenated with the perturbed image. The timestep sequence \(\{t_i\}\) can be sampled linearly or logarithmically.

This strategy injects explicit structural information through the conditional image, making the sampling process highly structure-aware and significantly reducing the number of sampling steps. For example, traditional generation from Gaussian noise often requires more than 100 steps. In contrast, our hijacking strategy reduces this to less than 20 steps while preserving or even improving the image quality.
Empirical results show that choosing \(t_0\) within 20\%-40\% of the noise schedule yields a good trade-off between convergence stability and sampling speed. The final output image can also be linearly fused with the optimized image from data consistency to further enhance perceptual quality and physical fidelity (see Subsection \ref{SubsectionRLF} for details).

\subsection{Physics-informed Data Consistency}

Although Poisson Flow-based generative models exhibit strong capabilities in structural completion and denoising for sparse-view CT reconstruction, it is important to recognize their potential limitations. When the projection angles are severely limited, the model may produce unrealistic textures or hallucinated structures. This issue arises from the fact that generative models primarily rely on data-driven mappings without explicit physical constraints. In scenarios with high information loss, the models tend to hallucinate non-existent tissues or edges, resulting in misleading reconstructions.
To overcome the problem of hallucinated structures in regions with missing information, we introduce a physics prior during the generation process by enforcing consistency constraints in the projection domain. Specifically, we explicitly incorporate physical operators (\emph {e.g.}, the projection operator and angular sampling mask) and combine them with data prior provided by ResPF. This dual prior fusion strategy has been shown in various studies to improve both the accuracy and stability of image reconstruction.

Within this framework, we adopt the efficient EDM sampler and embed the Adaptive Steepest Descent - Projection Onto Convex Sets (ASD-POCS) algorithm as a post-processing module during the generation stage. ASD-POCS applies the Total Variation (TV) minimization and non-negativity constraints to iteratively refine the generated image, ensuring consistency with the actual measured sinogram. The data consistency correction using ASD-POCS can be formulated as the following inequality-constrained variational optimization problem
\begin{equation}
\hat{\mathbf{x}} = \arg\min_{\mathbf{x}} \|\mathbf{x}\|_\text{TV},
\end{equation}
subject to the inequality constraint for data fidelity
\begin{equation}
\|\mathcal{M}(\Lambda)\mathcal{A} \mathbf{x} - \mathbf{y}_\text{sp}\|_2 \leq \epsilon,
\end{equation}
and the non-negativity constraint
\begin{equation}
\mathbf{x} \geq 0.
\end{equation}
Here, the TV norm is defined as
\begin{align}
\|\mathbf{x}\|_\text{TV} = \sum_{s,m} \big( &
(x_{s,m} - x_{s-1,m})^2 + \notag \\
& (x_{s,m} - x_{s,m-1})^2 \big)^{1/2},
\end{align}
where \(s, m\) denote the spatial indices within the image.

The ASD-POCS algorithm incorporates a key step to enhance data consistency, with the projection operation enforcing non-negativity. Notably, ASD-POCS minimizes the TV norm in an iterative fashion, guiding the reconstruction toward a TV-reduced solution. In each iteration, the algorithm alternates between the following two steps.\\
\noindent{\emph{Step 1: Steepest Descent Step (ASD)}}
\begin{equation}
    \mathbf{x}^{(k+\frac{1}{2})} = \mathbf{x}^{(k)} - \eta_k \cdot \nabla \|\mathbf{x}^{(k)}\|_\text{TV},
\end{equation}
where \(\eta_k\) is an adaptive step size to reduce artifacts and noise in the image.

\noindent{\emph{Step 2: Projection Onto Convex Sets (POCS)}}
\begin{equation}
    \mathbf{x}^{(k+1)} = P_C\left( \mathbf{x}^{(k+\frac{1}{2})} + \mu_k \mathcal{A}^\top \left( \mathbf{y}_\text{sp} - \mathcal{M}(\Lambda)\mathcal{A} \mathbf{x}^{(k+\frac{1}{2})} \right) \right),
\end{equation}
where \(\mu_k\) controls the update strength, and \(P_C(\cdot)\) denotes projection onto the feasible set (\emph{e.g.}, enforcing non-negativity).

After obtaining the initial reconstruction \(\hat{\mathbf{x}}\) from cPFGM sampling, we perform \(N_{dc} = 10\) iterations of ASD-POCS to obtain the final reconstruction. This post-processing module serves as a physics-informed consistency mechanism beyond the generative model, compensating for structural deviations in extremely sparse scenarios.

\subsection{Residual Linear Fusion}
\label{SubsectionRLF}
While the cPFGM is capable of generating structurally complete and perceptually natural image priors, its output \(\hat{\mathbf{x}}_{\text{gen}}\) may exhibit some degree of physical inconsistency in sparsely observed projection regions. On the other hand, the image refined via ASD-POCS post-processing, denoted as \(\hat{\mathbf{x}}_{\text{phys}}\), maintains good data fidelity but may suffer from oversmoothing and loss of fine details.
To combine the complementary strengths of these two approaches, as shown in Fig.\ref{fig:flow chart}(C) we propose a Residual Linear Fusion Strategy that linearly combines the generative output and the physically consistent reconstruction to produce a more balanced final result. The fused image is formulated as:
\begin{equation}
\hat{\mathbf{x}}_{\text{fused}} = \alpha \cdot \hat{\mathbf{x}}_{\text{gen}} + (1 - \alpha) \cdot \hat{\mathbf{x}}_{\text{phys}},
\end{equation}
where \(\hat{\mathbf{x}}_{\text{gen}}\) is the initial image generated by cPFGM, \(\hat{\mathbf{x}}_{\text{phys}}\)is the image refined by ASD-POCS post-processing, and \(\alpha \in [0,1]\) is the fusion coefficient that balances between generative prior and physical consistency.

Importantly, this residual fusion strategy not only provides a perceptual-physical trade-off at the image level, but also effectively addresses a key limitation of step-wise data consistency insertion in ODE-based sampling. In conventional approaches, repeatedly applying physical corrections at each sampling step often disrupts the smooth trajectory of the generative ODE, leading to unstable convergence and degraded image quality. By contrast, our fusion-based strategy decouples data fidelity from the intermediate steps and applies it at a controlled global level, preserving the continuity of the ODE sampling path while ensuring that the final result remains physically plausible.

\begin{figure}
    \centering
    \includegraphics[width=1\linewidth]{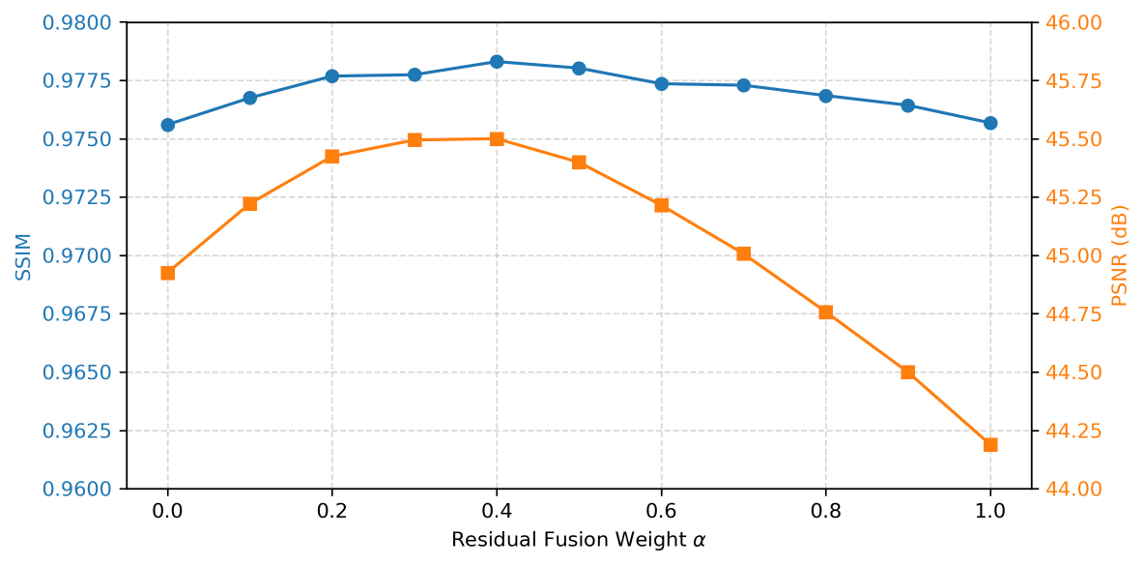}
    \caption{Effect of fusion coefficient $\alpha$ on SSIM and PSNR.}
    \label{fig:fusion weight}
\end{figure}

When \(\alpha\) is large, the result favors the generative output, yielding more natural and fine-grained structures. When \(\alpha\) is small, the result emphasizes physical consistency, enhancing fidelity to the observed measurements. Through a grid search over $\alpha \in [0, 1]$ conducted on the simulation dataset, we find that $\alpha = 0.4$  provides an effective trade-off between perceptual quality and data fidelity in sparse-view CT reconstruction. This value may depend on datasets and/or clinical scenarios. The corresponding quantitative metrics are presented in Fig.~\ref{fig:fusion weight}.

\begin{algorithm}[!b]
\caption{Sampling Procedure of ResPF}
\label{alg:respf_sample}
\begin{algorithmic}[1]
\REQUIRE Sparse sinogram $\mathbf{y}_{\text{sp}}$, projection operator $\mathcal{A}$, sampling mask $\mathcal{M}(\Lambda)$, noise schedule $\{t_i\}_{i=\tau}^T$, model $f_\theta$, fusion coefficient $\alpha$, number of ASD-POCS steps $N_{dc}$
\ENSURE Final fused reconstruction $\hat{\mathbf{x}}_{\text{fused}}$

\STATE $\mathbf{x}_{\text{sp}} \leftarrow \texttt{FBP}(\mathbf{y}_{\text{sp}})$ \hfill // Conditional image reconstruction
\STATE $\mathbf{x}_\tau \leftarrow \mathbf{x}_{\text{sp}}$ \hfill // Hijack at timestep $\tau$

\FOR{$i = \tau$ to $T-1$}
    \STATE $\mathbf{d}_i \leftarrow \frac{\mathbf{x}_i - f_\theta(\mathbf{x}_i, t_i, \mathbf{x}_{\text{sp}})}{t_i}$ \hfill
    \STATE $\tilde{\mathbf{x}}_{i+1} \leftarrow \mathbf{x}_i + (t_{i+1} - t_i) \cdot \mathbf{d}_i$
    \IF{$t_{i+1} > 0$}
        \STATE $\mathbf{d}_i' \leftarrow \frac{\tilde{\mathbf{x}}_{i+1} - f_\theta(\tilde{\mathbf{x}}_{i+1}, t_{i+1}, \mathbf{x}_{\text{sp}})}{t_{i+1}}$
        \STATE $\tilde{\mathbf{x}}_{i+1} \leftarrow \mathbf{x}_i + \frac{1}{2}(t_{i+1} - t_i)(\mathbf{d}_i + \mathbf{d}_i')$ \hfill 
    \ENDIF

    \STATE $\hat{\mathbf{x}}_{\text{gen}} \leftarrow \tilde{\mathbf{x}}_{i+1}$ \hfill // Generative output
    \STATE $\hat{\mathbf{x}}_{\text{phys}} \leftarrow \hat{\mathbf{x}}_{\text{gen}}$ \hfill // Data consistency initialization

    \FOR{$k = 1$ to $N_{dc}$}
        \STATE $\hat{\mathbf{x}}_{\text{phys}} \leftarrow \texttt{ASD-POCS}(\hat{\mathbf{x}}_{\text{phys}}, \mathbf{y}_{\text{sp}}, \mathcal{A}, \mathcal{M}(\Lambda))$
    \ENDFOR

    \STATE $\hat{\mathbf{x}}_{\text{fused}} \leftarrow \alpha \cdot \hat{\mathbf{x}}_{\text{phys}} + (1 - \alpha) \cdot \hat{\mathbf{x}}_{\text{gen}}$ 
    \STATE $\mathbf{x}_{i+1} \leftarrow \hat{\mathbf{x}}_{\text{fused}}$ \hfill // Prepare for next step
\ENDFOR

\RETURN $\hat{\mathbf{x}}_{\text{fused}}$
\end{algorithmic}
\end{algorithm}


\section{Experiments and Results}
\subsection{Experimental Setup}

To comprehensively evaluate the performance of the proposed ResPF in sparse-view CT reconstruction, experiments are conducted on two simulated datasets and one clinical dataset. The simulated datasets are used for performance validation and hyperparameter analysis, while the clinical dataset is used to evaluate the model's generalizability and practicality.

\subsubsection{Datasets}
\paragraph{AAPM Metal Artifact Challenge Dataset (Simulation)}
This dataset is used to investigate the impact of the dimensional expansion parameter $D$ in the PFGM++ framework. CT images are treated as numerical phantoms, the Siddon algorithm is used to generate sinograms assuming fan-beam geometry with $1000$ views and $900$ detector cells. Ground-truth images are reconstructed from full-view sinograms. Sparse-view sinograms are generated by uniformly sampling $77$, $125$, and $155$ views, and used to evaluate different dimensional settings $D \in \{64, 128, \infty\}$.

\paragraph{AAPM Low-Dose CT Challenge Dataset (Simulation)}
The Mayo Clinic dataset for the AAPM Low-Dose CT Grand Challenge \cite{aapm_low_dose_ct_grand_challenge} is used to simulate sparse-view CT data. This dataset contains 10 patients’ CT images reconstructed on a 512 × 512 grid with a slice thickness of 1 mm and a D30 kernel (medium). The physical size of each slice is $250\times 250$ $mm^{2}$, leading to a pixel size of about $0.781 \times 0.781$ $mm^2$. Using those image slices as phantoms, we numerically generate full-view sinograms via Siddon’s ray-driven algorithm \cite{siddon1985fast} with a typical fan-beam CT configuration (source-to-center: 550 mm, center-to-detector: 400 mm, detector interval  about 1/512 $rad$, 512 elements with equal-angle).  We simulate fan-beam projections with $1000$ views and $900$ detector cells to obtain full-view sinograms, from which reference ground-truth images are reconstructed. Sparse-view sinograms are obtained by uniformly sampling 63 and 125 views from the full-view sinograms, and sparse-view images are reconstructed using FBP. For training, we use 5,410 slices from 9 patients, reserving 526 slices from patient L506 for validation. This simulation framework ensures controlled evaluation of sparse-view CT reconstruction performance.

\paragraph{GE Clinical CT Dataset (Real)}
To further evaluate the generalizability of our proposed method under real clinical conditions, we apply the algorithm to a de-identified cardiac CT dataset acquired on a GE Discovery HD750 CT scanner. This dataset was obtained with approval from the Institutional Review Board  (IRB) at Vanderbilt University Medical Center, and the study was approved by the IRB at the University of Massachusetts Lowell. The patient was scanned in axial mode, and a total of 1,520 projections were acquired over a 556° angular range (approximately 1.54 rotations). Each projection consists of 835 detector elements with a pitch of 1.095 mm. The source-to-rotation-center distance is 538.5 mm, and the source-to-detector distance is 946.7 mm. Reconstruction is performed using a FBP algorithm with equiangular geometry, and the resulting image resolution is set to 512×512 to enable efficient sampling. The full-view projection data are collected from 984 views, we extract 123 views data from the full-view projection to evaluate the performance of algorithms.

\subsubsection{Baseline Methods}
We compare our method against seven representative deep learning-based CT reconstruction baselines, including convolutional image restoration networks, transformer-based models, and diffusion-based generative approaches.

\vspace{1mm}
\begin{itemize}
    \item FBPConvNet\cite{jin2017deep}: A U-Net based model that learns to refine FBP reconstructions through residual mapping.
    \item RED-CNN\cite{chen2017low}: A residual encoder-decoder network designed for denoising low-dose  CT images.
    \item SwinIR\cite{liang2021swinir}: A Swin Transformer-based image restoration model capable of capturing long-range dependencies.
    \item DDS\cite{chung2023decomposed}: A score-based generative model that iteratively reconstructs CT images through the diffusion process.
    \item PPFM\cite{hein2024ppfm}: PPFM is the first model to apply the Poisson flow generative model in CT reconstruction, which was designed for low-dose CT reconstruction tasks.
    \item EDM\cite{karras2022elucidating}:  EDM (Elucidated Diffusion Model) includes a family of diffusion models, and Poisson flow generative models are based on it, which offers fast sampling via denoising optimization.     
    As the augmented dimension $D \to \infty$, the PFGM converges to EDM, and therefore, the improvements in this work are equally applicable to EDM. Experimental results show that PFGM achieves satisfactory reconstruction quality with as few as 16 sampling steps, whereas EDM typically requires 32 steps or more to reach comparable performance. To ensure a fair comparison and to evaluate the effectiveness of the proposed strategies, EDM is tested using the same configuration and sampling parameters as ResPF throughout this study.
\end{itemize}

All the baseline methods are retrained from scratch on the same training set as our method. Inputs and outputs are aligned in size and format, and all methods are evaluated under identical simulation and clinical testing protocols.

\subsubsection{Implementation Details}

\begin{figure}
    \centering
    \includegraphics[width=1\linewidth]{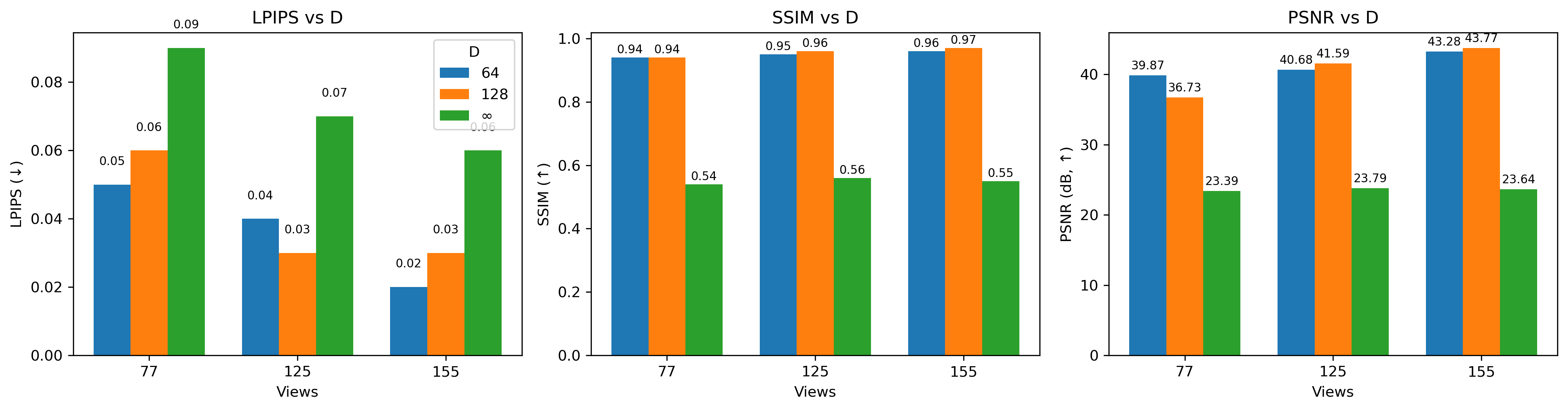}
    \caption{Effect of $D$ for reconstruction under different sparse-view projections.}
    \label{fig:D effect}
\end{figure}
Our method is implemented based on the PFGM++ framework, with DDPM++ backbone and EDM as the base sampler. The model is trained using the Adam optimizer with a learning rate of $2\times10^{-4}$, batch size of 16, and 3 million training iterations. During training, random $256 \times 256$ patches are cropped from the input images for data augmentation and memory efficiency. The network uses a channel multiplier of 128 with channel configuration $[1, 1, 2, 2, 2, 2, 2]$, and self-attention layers are inserted at resolutions 16, 8, and 4. Preconditioning, exponential moving average (EMA), and 15\% random augmentations are also applied.
During sampling, we use 16 steps by default within the EDM framework, 2 hijcak steps are incorporated to accelerate convergence. We also explore an optional data consistency module (ASD-POCS) and residual linear fusion with traditional reconstructions. These components are further analyzed in the ablation study section.

In PFGM++, the dimensional parameter $D$ determines the expansion of data into a higher-dimensional Poisson field. Systematic experiments are conducted on the AAPM Metal Artifact Challenge simulations dataset with $D \in \{64, 128, \infty\}$, and the results are showed in Fig.\ref{fig:D effect}. It can be seen that $D = 64$ and $D = 128$ yield similar reconstruction quality. However, $D = 128$ offers better numerical stability and convergence behavior. Thus, we adopt $D = 128$ as the default setting.
Different sampling steps are also tested in $\{8, 16, 32\}$ and it is found that 16 steps offer a good trade-off between reconstruction quality and computational efficiency. Therefore, all experiments are conducted using 16 sampling steps.

In the data consistency step, we adopt the ASD-POCS algorithm to refine the generated image toward better alignment with the measured projections. Benefiting from the fact that the initial input for ASD-POCS originates from the generative result of the previous sampling step, the iterative process converges more rapidly compared to conventional settings. Consequently, we configure the algorithm with 10 iterations and 8 subsets, and apply 5 total variation (TV) minimization steps after each iteration to enhance edge preservation and noise suppression.

\subsubsection{Evaluation Metrics}

We evaluate reconstructed image quality using three widely adopted metrics. (a) PSNR (Peak Signal-to-Noise Ratio) measures overall reconstruction accuracy. Higher PSNR indicates better fidelity. 
(b) SSIM (Structural Similarity) measures luminance, contrast, and structural similarity. More sensitive to structural distortions.
(c) LPIPS (Learned Perceptual Image Patch Similarity) is a perceptual metric based on deep feature distances from AlexNet, and more aligned with human visual perception. It is the primary metric in this study. 
For simulated datasets, all metrics are computed using image phantom as reference. For the GE clinical dataset, evaluations are based on reconstructions from full-view (984-view) sinograms.

\subsection{Numerical Simulations Results}

\begin{figure*}[t]
    \centering
    \includegraphics[width=1\linewidth]{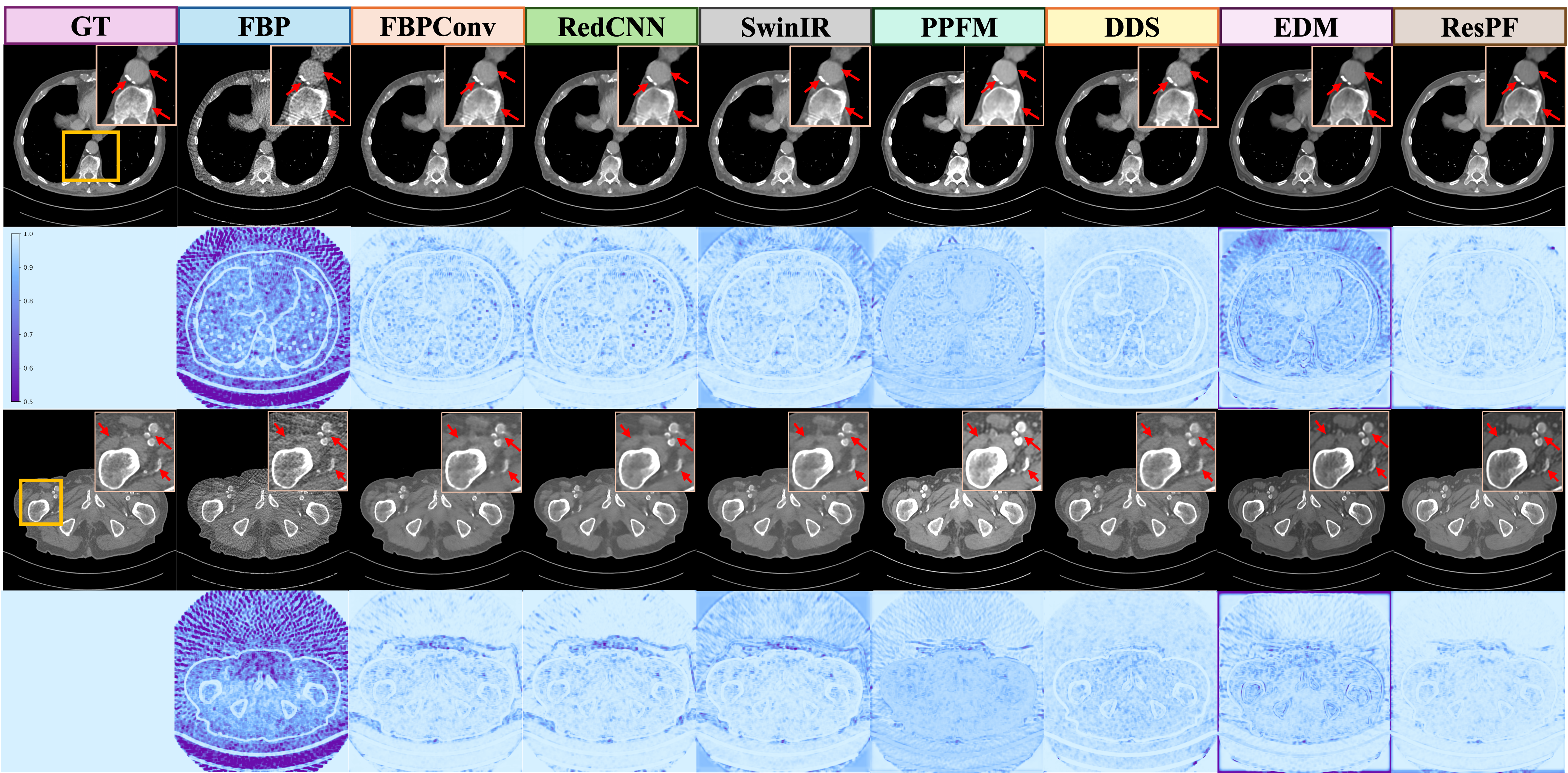}
    \caption{Visual comparison of reconstruction results on two slices from the AAPM low-dose CT simulation dataset under a 125-view setting.
The first and third rows show reconstructed images in a display window [–400, 500]HU, while the second and fourth rows present SSIM maps with respect to the ground truth. Zoomed-in regions of interest (ROIs) highlight anatomically fine structures such as vessels and soft tissues.
}
    \label{fig:125}
\end{figure*}

\begin{figure*}[t]
    \centering
    \includegraphics[width=1\linewidth]{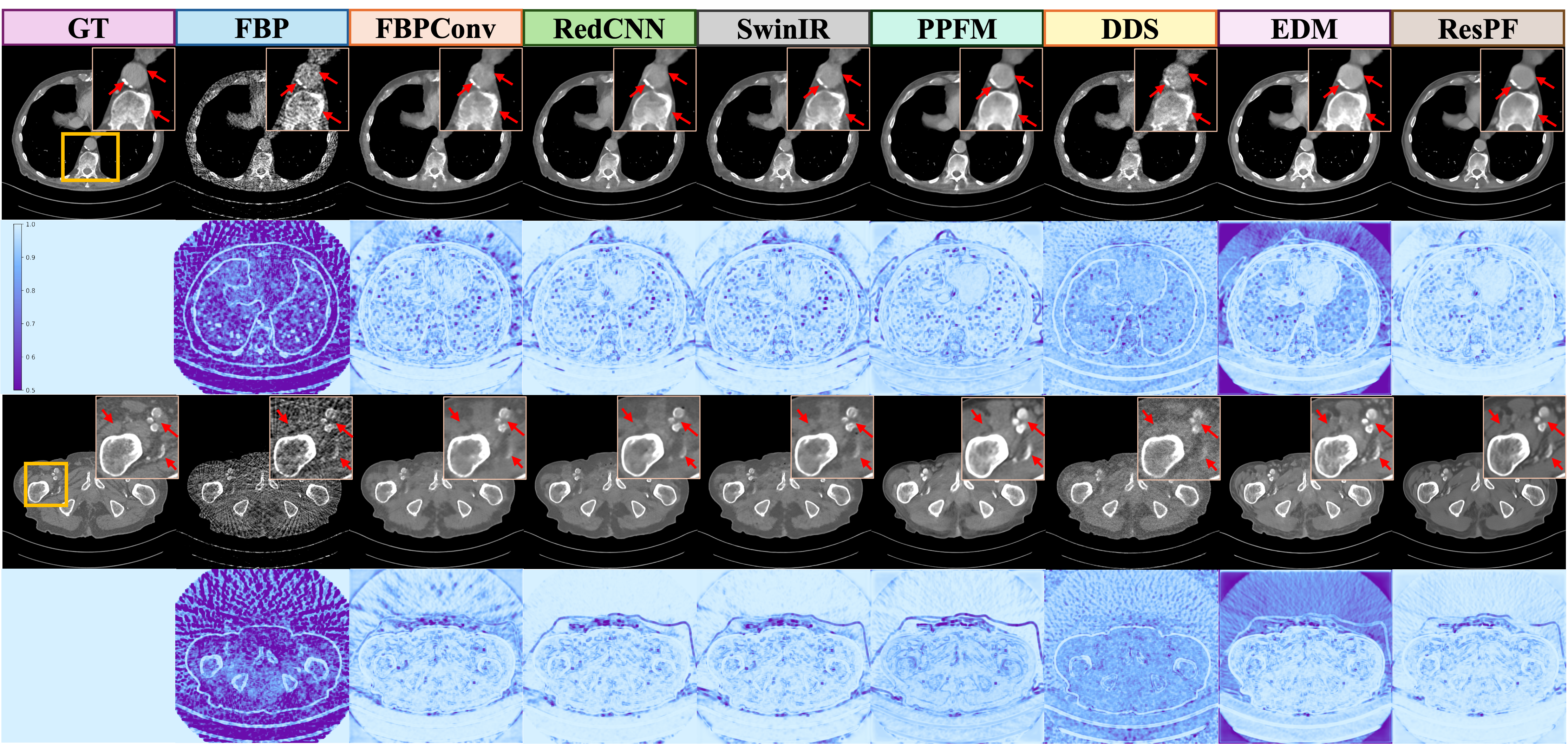}
    \caption{Same as Fig.\ref{fig:125} but reconstructed from highly sparse 63-view setting.}
    \label{fig:63}
\end{figure*}

To qualitatively evaluate ResPF performance under different sparse levels, Figs.\ref{fig:125} and \ref{fig:63} present visual comparisons for two representative anatomical slices from the AAPM low-dose CT simulation dataset under 125- and 63-view settings, respectively. Each subfigure includes reconstructed images and the corresponding SSIM maps, with highlighted regions of interest (ROIs) to emphasize diagnostically critical areas particularly sensitive to undersampling artifacts.

In the case of 125-view (Fig.\ref{fig:125}), CNN-based methods, such as FBPConvNet and REDCNN, exhibit moderate noise suppression but suffer from noticeable blurring and structural distortions, especially around vessel boundaries and soft tissue interfaces. Although SwinIR and DDS achieve better texture preservation, their reconstructions remain susceptible to residual artifacts and localized intensity inconsistencies. EDM tends to generate excessively smooth textures and slightly distorted edge features in underdetermined regions. In contrast, ResPF consistently delivers the most structurally coherent reconstructions, characterized by sharper boundaries and superior local fidelity within the ROIs. The corresponding SSIM maps confirm this observation, highlighting the uniform and elevated similarity achieved by ResPF.

In the case of more sparse 63-view (Fig.\ref{fig:63}), performance differences become even more obvious. Classical deep learning approaches fail to adequately recover fine anatomical details, displaying prominent streak artifacts. Even advanced deep learning methods begin to hallucinate structures or experience detail collapse in highly undersampled regions. For example, EDM demonstrates decreased robustness under extreme sparsity, necessitating longer sampling trajectories to yield competitive reconstructions. In contrast, ResPF maintains stable and coherent reconstruction quality with minimal streaking and artifact propagation, benefiting from the residual fusion mechanism that effectively balances generative priors with physical consistency. SSIM maps further confirm ResPF’s consistent retention of higher local similarity, particularly within complex anatomical areas, highlighting its capability to preserve fine structures despite aggressive undersampling.

Quantitative evaluations further substantiate these findings, as summarized in Table \ref{tab:metrics table}. In the case of severely undersampled 63-view, the ResPF framework consistently achieves the best performance across all metrics, demonstrating significant advantages in both structural fidelity and perceptual quality. EDM underperforms in this low-data regime, as indicated by a notable reduction in SSIM (0.8754) and elevated LPIPS (0.0908), highlighting its limited robustness under extreme undersampling.
In the case of less constrained 125-view, all methods show improved reconstruction quality, yet ResPF continues to outperform competing approaches, reinforcing the efficacy of our residual fusion strategy across different sparse levels.


Overall, these results clearly illustrate that ResPF achieves state-of-the-art reconstruction performance in both structural accuracy and perceptual fidelity while maintaining robustness in different sparse scenarios, positioning it as a highly promising method for practical sparse-view CT reconstruction applications.


\begin{table*}[t]
\centering
\caption{Quantitative evaluation results on simulation and clinical datasets. \textbf{Bold: best}, \underline{Underscore: Second best.}}
\label{tab:metrics table}
\resizebox{\textwidth}{!}{%
\begin{tabular}{cccccccccc}
\hline
\multirow{3}{*}{Method} & \multicolumn{6}{c}{Simulation Dataset}                                 & \multicolumn{3}{c}{Clinical Dataset} \\ \cline{2-7} \cline{8-10} 
                        & \multicolumn{3}{c}{63 views} & \multicolumn{3}{c}{125 views}        & \multicolumn{3}{c}{123 views}      \\ \cline{2-7} \cline{8-10}
                        & SSIM ↑  & PSNR ↑   & LPIPS ↓ & SSIM ↑           & PSNR ↑  & LPIPS ↓ & SSIM ↑    & PSNR ↑    & LPIPS ↓   \\ \hline
FBPConvNet\cite{jin2017deep}              & 0.9379  & 37.1205  & 0.0720  & 0.9688           & 40.5960 & 0.0380  &            0.8090&           31.6507&            0.1916\\
SwinIR\cite{liang2021swinir}         & \underline{0.9528} & \underline{37.9800}& \underline{0.0660} & 
\underline{0.9697}& \underline{41.7400} & \underline{0.0362} &  0.7985&  32.1689&  \underline{0.1527}\\
RED-CNN\cite{chen2017low}                 & 0.9379  & 37.1655  & 0.0718  & 0.9520& 40.9736 & 0.0377  &           0.8209&           32.4415&           0.1848\\
DDS\cite{chung2023decomposed}                     & 0.9394  & 37.1263  & 0.0725  & 0.9658           & 39.5536 & 0.0406  &     0.8019      &    31.5209       &    0.2273       \\
PPFM\cite{hein2024ppfm}                    & 0.9346  & 30.7244  & 0.0896  & 0.9411           & 28.5001 & 0.0518  &           0.8165&           31.4364&           0.2502\\
EDM (D = $\infty$)\cite{karras2022elucidating}      & 0.8754  & 38.5046  & 0.0908  & 0.9541           & 31.1091 & 0.0412  &           \underline{0.8298}&           \underline{33.3157}&           0.1593\\
Ours (D = 128) & \textbf{0.9639}  & \textbf{39.8028}& \textbf{0.0630}  & \textbf{0.9796} & \textbf{45.1912}& \textbf{0.0333}&  \textbf{0.8581}& \textbf{34.2266}& \textbf{0.1124}\\ \hline
\end{tabular}%
}
\end{table*}

\subsection{Results on Clinical Data}

\begin{figure*}[t]
    \centering
    \includegraphics[width=1\linewidth]{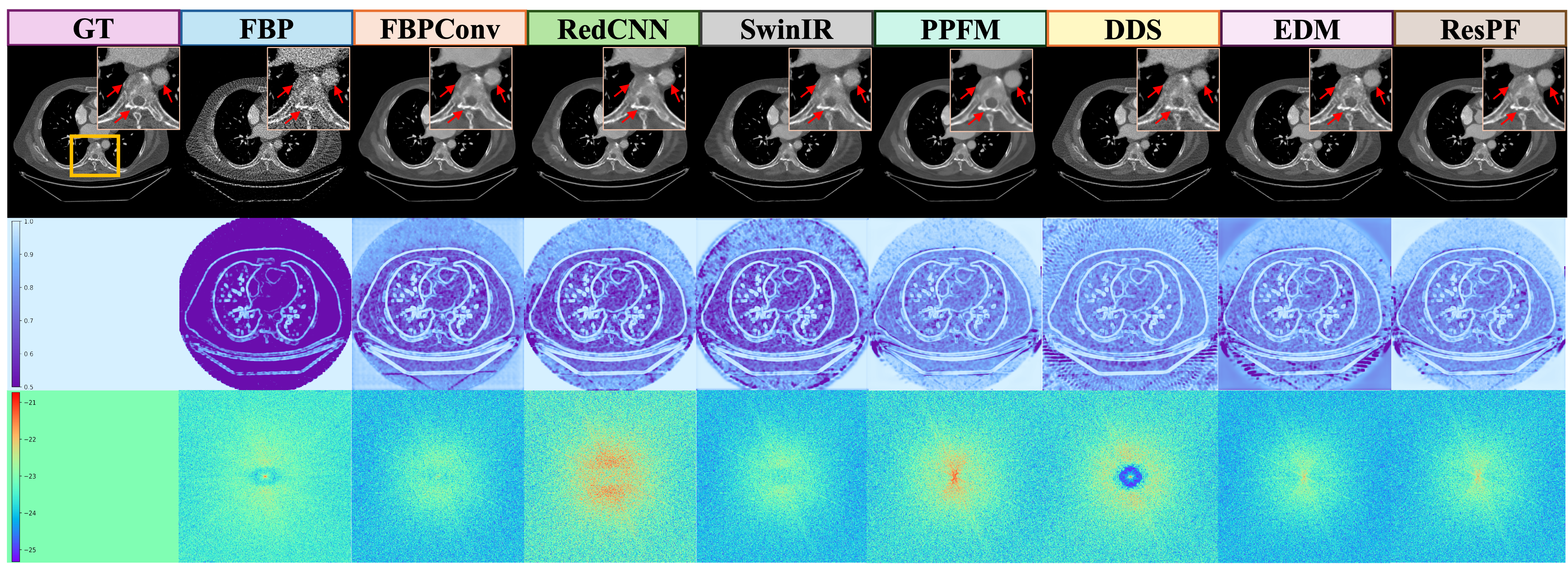}
    \caption{Qualitative and frequency-domain comparison of reconstruction methods on clinical CT dataset under a 123-view sparse-view setting.
The first row shows reconstructed images with a display window of [–450, 880] HU, with zoom-in ROIs in the upper-right corners highlighting vascular boundaries and lung textures. The second row presents SSIM maps with respect to the clinical reference, indicating local structural similarity. The third row shows Noise Power Spectrum (NPS) maps that reveal frequency-domain artifacts and texture degradation.
}
    \label{fig:GE data}
\end{figure*}
To validate the generalizability of the proposed ResPF framework, we further evaluate it on clinical sparse-view CT data. Cardiac CT projections are acquired on a GE CT scanner in a circular scan geometry. 984 projections are acquired in a full scan protocol. To simulate realistic sparse-view data acquisition, we purposely discard 7 in every 8 projections, resulting in 123 projection views. Fig.~\ref{fig:GE data} presents a qualitative comparison for several representative methods, including traditional, deep learning, and generative approaches. Ground-truth image is reconstructed from the original full scan projections using FBP. 
From the reconstructed images (first row), FBP and FBPConvNet exhibit strong streaking artifacts and loss of anatomical detail, especially along lung boundaries. RED-CNN and SwinIR improve smoothness but over-smooth fine structures. Among the generative models, EDM and DDS produce visually natural textures but introduce geometric distortions or detail hallucination in homogeneous areas. PPFM suppresses noise effectively but over-smooths high-frequency content, leading to texture loss. The SSIM maps (second row) reveal local structural similarity. CNN-based methods show low SSIM near edges, while EDM and DDS perform better in flat regions but degrade near complex boundaries. ResPF maintains high SSIM consistently, especially around key anatomical features, indicating better structural preservation. The NPS maps (third row) provide a frequency-domain view of residual noise. DDS shows directional bias with concentrated energy, while PPFM overly attenuates high-frequency components. EDM displays uneven frequency retention. In contrast, ResPF exhibits a smooth and isotropic NPS distribution, indicating natural texture recovery without introducing excessive noise. These results align with visual and SSIM findings, confirming ResPF’s overall advantages in spatial accuracy and noise realism.

Table~\ref{tab:metrics table} reports the quantitative evaluation of the clinical dataset. Overall, ResPF outperforms all baseline methods in all indicators, demonstrating superior perceptual quality, structural fidelity, and data consistency. Although traditional learning-based methods achieve competitive results under moderate undersampling, they struggle to generalize under complex anatomical variability. Generative models such as PPFM, DDS, and EDM show improved image quality but often fail to balance fine-detail recovery with data fidelity. ResPF achieves the best trade-off among these factors, highlighting the effectiveness of the residual fusion strategy in real-world clinical scenarios.


\subsection{Runtime Efficiency Comparison}

Beyond reconstructed image quality, runtime efficiency is a critical factor for the practical deployment of generative models in clinical workflows. Table~\ref{tab:sampling time}  reports the average sampling time per image for each generative model, evaluated on both simulated and clinical datasets. DDS, which relies on high-resolution DDIM-style sampling with approximately 50 deterministic SDE steps, exhibits significantly longer runtimes than all other methods. This computational burden limits its suitability for time-sensitive clinical applications, despite its reconstruction quality. In contrast, PPFM, based on the Poisson flow framework with a deterministic ODE sampling path, supports single-step generation, achieving the fastest runtime. However, as previously discussed, this speed comes at the cost of reduced structural fidelity in challenging anatomical regions.

ResPF achieves a favorable balance, offering sampling efficiency comparable to that of PPFM while delivering substantially better reconstruction quality. This efficiency is primarily attributed to two key strategies: (1) a residual fusion mechanism that enforces data consistency without interrupting the sampling flow, and (2) a hijacking strategy that bypasses the initial noise-dominated stages and reduces the total number of ODE steps.
Together, these techniques make ResPF well-suited for real-time or near-real-time CT reconstruction scenarios.

Although EDM achieves a similar sampling time, it produces lower-quality reconstructions compared to ResPF. This discrepancy can be attributed to the fundamental difference between the two approaches. While ResPF is based on the Poisson flow formulation, EDM is rooted in diffusion models. As a result, despite sharing compatible sampling strategies, ResPF benefits from a more expressive and geometry-aligned generative prior, leading to superior reconstruction performance.
\begin{table}[]
\centering
\caption{Sampling time comparison for generative models}
\label{tab:sampling time}
{%
\begin{tabular}{ccc}
\hline
\multirow{2}{*}{Method} & \multicolumn{2}{c}{Time per Image (s)} \\ \cline{2-3} 
                                 & Simulation Data  & Clinical Data  \\ \hline
DDS\cite{chung2023decomposed}                              & $27.15 \pm 0.12 $                     &                  $39.26 \pm 0.37  $    \\
PPFM\cite{hein2024ppfm}                             & $ 0.53 \pm 0.05  $                      &                     $0.56\pm 0.05 $\\ 
EDM\cite{karras2022elucidating}                              & $ 1.55 \pm 0.08  $                      &                     $2.02\pm 0.13  $\\
ResPF(Ours)                      & $1.54 \pm 0.08  $                     &                     $1.72 \pm 0.16  $\\  \hline
\end{tabular}%
}
\end{table}

\subsection{Ablation Study}
\begin{figure}[t] 
    \centering
    \includegraphics[width=1\linewidth]{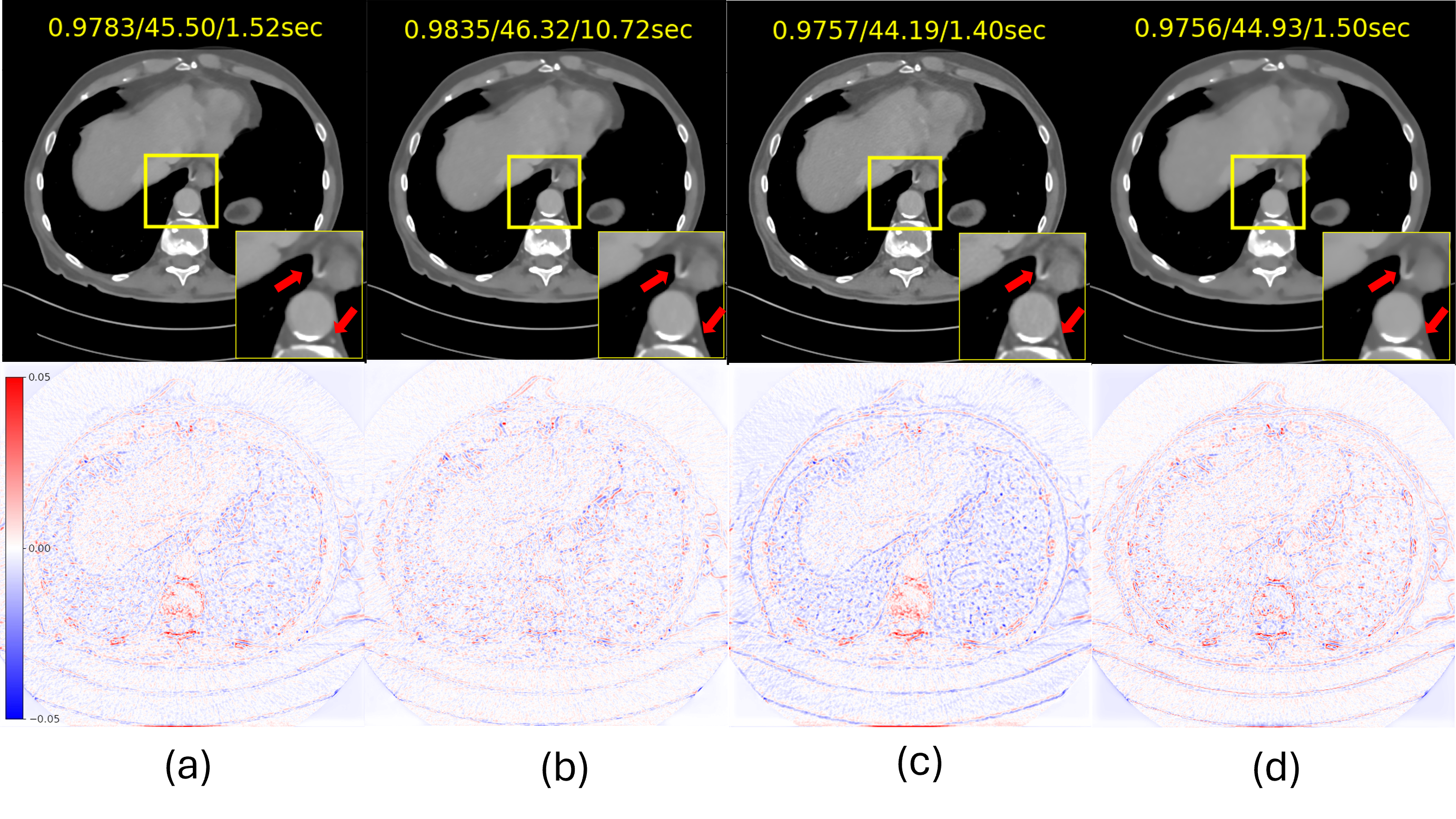}
    \caption{Ablation study results under 125-view sparse setting. Each column shows a different model configuration: (a) full ResPF with hijacking, data consistency, and residual fusion; (b) without hijacking; (c) without data consistency; and (d) without residual fusion only. For each setup, the reconstructed image (top) in a display window [-400, 500] HU and the corresponding signed residual map (bottom) are shown.
    Although removing hijacking (b) achieves slightly better image quality (SSIM: 0.9835, PSNR: 46.32), it significantly increases sampling time (10.72s) compared to the full model (a), which offers comparable quality (SSIM: 0.9773, PSNR: 44.95) in just 1.52s. This highlights the trade-off between quality and efficiency, and underscores the practical benefit of our accelerated ResPF design.}
    \label{fig:ablation}
\end{figure}
To investigate the contribution of each key component in our proposed framework, we conduct ablation studies on the simulated dataset with 125 views. We focus on evaluating the following three modules: Hijack sampling strategy, data consistency module, and residual fusion module.
As shown in Fig.~\ref{fig:ablation}, four configurations are compared, 
and SSIM, PSNR, and inference time are annotated in the top-left corner of each reconstructed image.

\subsubsection{Full Model (a)}
The full model (configuration a) achieves competitive reconstruction quality, accurately recovering anatomical boundaries and fine textures, with only marginal differences compared to the best-performing configuration. The residual map shows minimal errors primarily located in flat regions, indicating structurally consistent sampling. More importantly, the reconstruction time remains extremely low, demonstrating that the hijack strategy enables highly efficient sampling while maintaining high image fidelity. This confirms that configuration (a) offers a favorable trade-off between reconstruction accuracy and inference speed, making it well-suited for time-critical CT applications.

\subsubsection{Without Hijack Strategy (b)}
In configuration (b), the model disables the hijack mechanism and performs the full 16-step conditional Poisson flow sampling process from random noise. As shown in Fig.~\ref{fig:ablation}, this configuration achieves the highest reconstruction quality among all variants, preserving sharp edges, natural gray-level transitions, and detailed anatomical structures. The residual map exhibits the lowest error magnitude, particularly in central and peripheral regions, indicating well-aligned sampling trajectories and accurate convergence toward the data manifold.

However, this performance comes at the cost of significantly increased sampling time. Without hijacking, the model must traverse the entire generative path from noise to data space, resulting in a per-image inference time of 10.72 seconds—an order of magnitude slower than our full model configuration (a), which completes sampling in just 1.52 seconds. While configuration (b) offers marginal gains in SSIM and PSNR, the runtime overhead limits its practicality, especially in clinical scenarios where fast reconstruction is critical.

In contrast, our full model adopts the hijacking strategy (starting from step 14 and sampling only 2 steps), which significantly reduces inference time while maintaining competitive reconstruction quality. This trade-off demonstrates that ResPF with hijack provides an effective balance between performance and efficiency, offering near-optimal quality with real-time capability.

\subsubsection{Without Data Consistency (c)}
Configuration (c) removes the data consistency module and directly outputs images from the cPFGM model. Although the SSIM and PSNR appear relatively high, the residual map reveals structural distortions and local artifacts. In the ROI region, clear misalignment is observed at the aorta boundary, and undesired texture smoothing artifacts emerge. This indicates that even with a powerful generative model, the absence of physics-informed constraints may lead to hallucinated structures, especially under sparse-view settings. Data consistency is thus crucial for suppressing reconstruction bias and ensuring physical plausibility.

\subsubsection{Without Residual Fusion (d)}

Configuration (d) removes the residual fusion between the sampling image and the data-consistent reconstruction (from ASD-POCS), and directly adopts the latter as the final output. Although hijack and data consistency are retained, the performance degrades significantly, clearly demonstrating the importance of residual fusion.
The reconstructed image shows noticeable ringing artifacts around structural edges, with subtle misalignments in vascular contours. The residual map indicates concentrated error near the aorta and surrounding tissues, suggesting that relying solely on ASD-POCS fails to preserve high-frequency details. While ASD-POCS enforces projection fidelity, its strong regularization or imperfect initialization may result in oversmoothing. In contrast, the sampling output from our conditional Poisson Flow generative model contains rich prior structures and high-frequency textures. Fusing both images allows complementary advantages: the sampling result preserves detail, and the data-consistent result ensures physical accuracy, achieving a better trade-off between perceptual quality and data fidelity. Removing this module leads to degraded structure integrity and overall consistency.

The three modules are complementary and jointly essential: the data consistency module enforces physical constraints to recover faithful structures; the residual fusion module enhances detail preservation and perceptual quality; the hijack strategy improves sampling efficiency and stability. Together, they enable accurate, physically consistent, and computationally efficient sparse-view CT reconstruction.

\section{Discussion and Conclusion}

This study for the first time presents a Residual Poisson Flow (ResPF) Generative Model for sparse-view CT image reconstruction, marking the first application of the PFGM++ framework in this domain. Leveraging the electrostatic formulation of PFGM++, ResPF reformulates the reconstruction task as learning a conditional joint distribution from sparse-view measurements to high-fidelity CT images. By modeling data distributions as electric charges in an augmented space of dimensionality $N + D$, the framework enables flexible adjustment of the augmented dimension to optimize the generative trajectory, facilitating accurate reconstruction under limited projection scenarios.

The key contributions of this work are threefold. First, a hijacking strategy is proposed to bypass early-stage noise sampling, substantially reducing the number of sampling steps and improving inference efficiency. Second, to mitigate the quality degradation caused by early diffusion skipping, a data consistency mechanism is incorporated at each sampling step, guiding the generation toward measured projections and enhancing physical fidelity. Third, to preserve the continuity of the ODE-based sampling process in the presence of data consistency updates, a residual fusion module is introduced. Inspired by ResNet architectures, it linearly combines generative outputs and data-consistent reconstructions, ensuring smooth and stable generative trajectories.

Experimental results on both simulated (125-view and 63-view) and real-world (123-view GE) datasets demonstrate the superior performance of ResPF. The model achieves consistently strong results across SSIM, PSNR, and LPIPS metrics, particularly under extreme undersampling conditions (\emph{e.g.}, 63 views), where it successfully recovers fine anatomical structures. Moreover, the integrated hijacking and residual fusion strategies significantly improve reconstruction efficiency without sacrificing quality. These findings collectively validate the effectiveness and practicality of ResPF for sparse-view CT reconstruction.

Despite ResPF’s promising results, several avenues remain for future work. While ASD-POCS ensures data consistency, alternative learned or optimization-based schemes could enhance speed and robustness. ResPF's reliance on manually tuned hyperparameters limits scalability; integrating automated optimization methods \cite{fang2022optimal,li2023hybrid} may improve efficiency and adaptability. A key extension is applying ResPF to 3D volumetric reconstruction, leveraging inter-slice continuity for better clinical utility. However, this introduces new challenges in memory usage, high-dimensional sampling stability, and enforcing 3D consistency, which warrant further investigation.

In conclusion, the proposed ResPF framework provides a fast, physically informed, and high-fidelity solution to sparse-view CT reconstruction. By combining conditional generative modeling, physics-guided correction, and trajectory-preserving fusion, ResPF offers a principled and scalable approach to improve reconstruction quality under challenging acquisition conditions. Future work will focus on extending the dataset diversity, refining the sampling and consistency modules, and advancing toward clinical translation.

\section*{Acknowledgement}
All authors declare that they have no known conflicts of interest in terms of competing financial interests or personal relationships that could have an influence or are relevant to the
work reported in this paper.

\bibliographystyle{IEEEtran}

\end{document}